\documentclass[12pt]{article}
\usepackage{epsfig}

\def\beq{\begin{equation}}
\def\eeq#1{\label{#1}\end{equation}}
\def\eeqn{\end{equation}}

\def\beqa{\begin{eqnarray}}
\def\eeqa#1{\label{#1}\end{eqnarray}}
\def\eeqan{\end{eqnarray}}

\let\bar=\overbar

\def\Dslash{\not{\hbox{\kern-4pt $D$}}}
\def\dslash{\not{\hbox{\kern-2pt $\del$}}}

\def\msb{{\bar{\ssstyle M \kern -1pt S}}}

\def\Title#1{\begin{center} {\Large {\bf #1} } \end{center}}

\begin{document}

\Title{Searches for Permanent Electric Dipole Moments}

\begin{center}{\large \bf Contribution to the proceedings of HQL06,\\
Munich, October 16th-20th 2006}\end{center}

\bigskip\bigskip

\begin{raggedright}  

{\it Klaus Jungmann\index{Jungmann, K.}\\
Kernfysisch Versneller Instituut\\
University of Groningen\\
NL-9747 AA Groningen, THE NETHERLANDS}
\bigskip\bigskip
\end{raggedright}

\section{Introduction}
The Standard Model (SM) \index{Standard Model}
of particle physics provides a theoretical framework
which allows to describe all observations in particle physics to date.
Even those recent observations in neutrino physics which strongly
indicate the existence of neutrino oscillations can be accommodated with
small modification. However,  contrary to this the great success of the SM
there remains a number of most intriguing questions in modern physics to which the
SM can not provide further clues about the underlying physical processes, 
although the facts are can be described often to very high accuracy.
Among those puzzling issue  are  the existence  of exactly 
three generations of fundamental particles, i.e. quarks and leptons, and the
hierarchy of the masses of these fundamental fermions.
In addition, the electro-weak SM has a rather large number of some 
27 free parameters, which all need to be extracted from experiments\cite{PDG_2006}.

In modern physics - and in particular in the SM -
symmetries play an important and central role. We know that
global symmetries relate to conservation laws and local symmetries 
give forces \cite{Lee_56}.
Within the SM
the physical origin of the observed breaking of discrete 
symmetries in weak interactions,
e.g. of parity (P), of time reversal (T) and of 
combined charge conjugation and parity (CP), 
remains unrevealed, although the experimental findings can be well
described.

In order to gain deeper insights into the not well understood 
aspects of fundamental physics, a number of
speculative models beyond the present standard theory
have been proposed. Those include such which involve Left-Right symmetry, 
fundamental fermion compositeness, new particles, leptoquarks, 
supersymmetry, supergravity, technicolor and many more. Interesting candidates 
for an all encompassing quantum field theory are string or membrane
(M) theories which among other features  may include supersymmetry in their low energy limit.
Independent of their mathematical elegance and partial appeal
all of these speculative theories will remain without status in physics
unless secure experimental evidence for them being reality 
can  be gained in future. Experimental searches
for predicted unique features of those models - such as breaking of 
discrete symmetries - are therefore essential
to steer the development of theory towards a better and deeper understanding of 
fundamental laws in nature. Such experiments must be carried out 
not only at high energy accelerators, but also in complementary approaches
at lower energies. Typically the low energy experiments in this context fall 
into the realm of atomic physics and of high precision measurements.
Searches for permanent Electric Dipole Moments (EDMs), \index{EDM} are an important
subset of such low energy precision particle physics experiments.

\section{Discrete Symmetries} \index{discrete symmetries}
A permanent electric dipole moment (EDM) \index{EDM} of any fundamental particle 
or quantum system violates both parity (P) and time reversal (T) 
symmetries \cite{Khriplovich_1997} and if one assumes unbroken combined
CPT symmetry, it also violates CP.
The violation of P is well established in physics and its 
accurate description has contributed significantly to the credibility of the SM.
The observation of neutral currents together with the
observation of parity non-conservation in atoms were
important to verify the validity of the SM. The fact that 
physics over several orders of magnitude  in momentum transfer 
- from atoms to highest energy scattering -
yields the same electro-weak parameters may be viewed as 
one of the biggest successes in physics to date.
However, at the level of highest precision electro-weak experiments
questions arose, which ultimately may call for a refinement.  

\subsection{Parity} \index{parity}
The  running of the weak mixing angle $sin^2 \Theta_W$
appears  not to be in good agreement with observations \cite{Czarnecki_1998}.
If the value of   $sin^2 \Theta_W$  is fixed at the Z$^0$-pole, deep inelastic electron scattering
at several GeV appears to yield a considerably higher value.
A new round of experiments is being started with the Q$_{weak}$ experiment \cite{Qweak} at the 
Jefferson Laboratory in the USA.  For atomic parity violation \cite{Haxton_2001} in principle
higher experimental accuracy will be possible from experiments using Fr
isotopes\cite{Atutov_2003,Gomez_2004} or single Ba or Ra ions in radio frequency traps 
\cite{Fortson}. Experiments with Fr atoms in magneto-optical traps 
were started at INFN Legnaro, Italy,  and at  Stony Brook, USA. Pioneering work on
single Ba ions has been started at the University of Washington, Seattle, USA, 
and based on this  experience  a Ra ion experiment has been  started at KVI, Groningen, The
Netherlands. The full exploitation of the advantage of larger weak effects in these heavy atom systems
(compared to Cs) due to their high power dependence on
the nuclear charge, will require improved atomic wave function
calculations, as the observation of weak effects is always through 
an interference of weak and electromagnetic effects.\cite{Sapirstein_2004}

\begin{figure}[tbh]
\begin{center}
\vspace*{1.9in}
\hspace*{-1cm} \epsfig{file=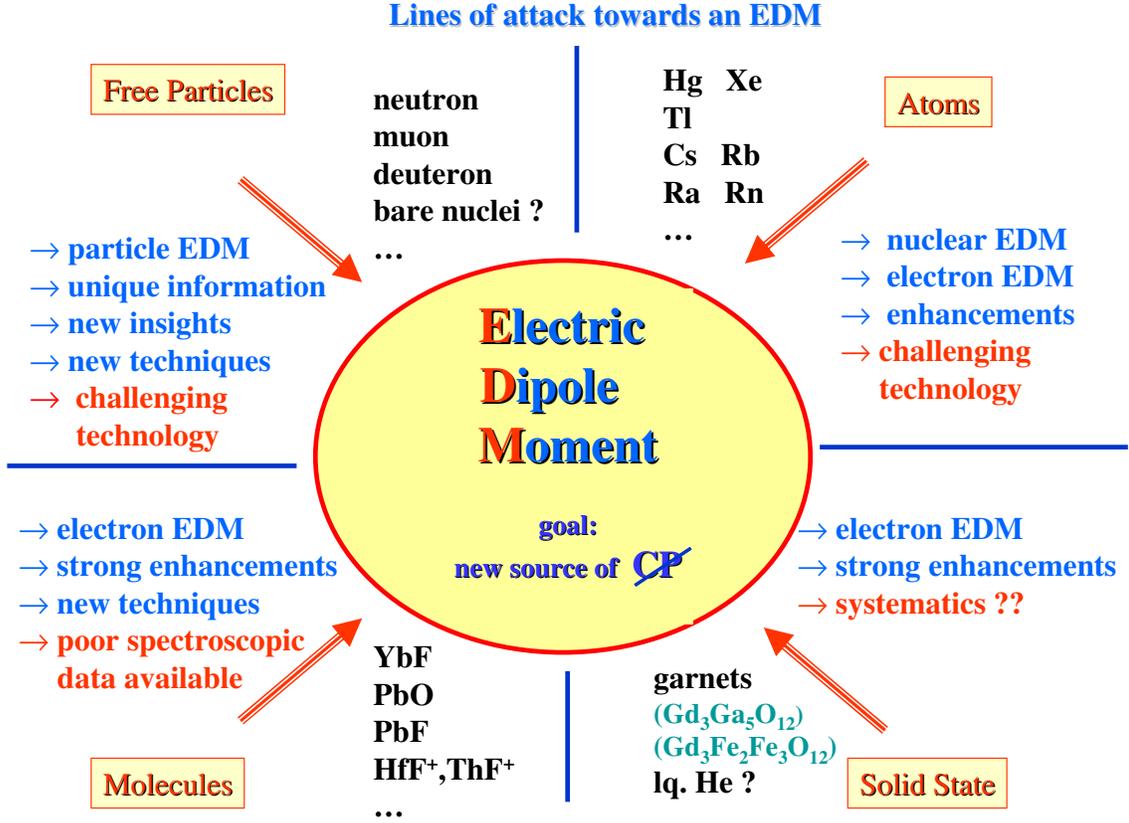,height = 4.4in}
\vspace*{-2in}   
\caption{ The experimental searches for an EDM
follow a number of significantly different lines. }
\label{fig_edm_attack}
\end{center}
\end{figure}

\subsection{Time Reversal and CP} \index{time reversal} \index{CP violation}
CP violation as observed first in the neutral Kaon decays 
can be described with a single phase factor in the Cabbibo-Kobayashi-Maskawa
formalism\cite{Cronin_1981}.
Because of its possible relation to the observed matter-antimatter 
asymmetry in the universe, CP violation has attracted a lot of attention. 
\footnote{ A. Sakharov\cite{Sakharov_1967} has suggested that the 
observed dominance of matter could be explained via CP-violation in the early universe 
in a state of thermal non-equilibrium and with baryon number violating
processes. We note here that the existence of additional sources of CP-Violation is
not a necessary condition to explain the matter-antimatter asymmetry.
Other viable routes could lead through CPT violation and there without the need
of thermal non-equilibrium\cite{Kostelecky_1996}.\index{ baryon asymmetry}}     
CP violation as described in the SM 
is however not sufficient to explain the excess of baryons.
This provides a  strong motivation to search for yet unknown sources 
CP symmetry violation. It is in particular  a major driving force behind
theEDM searches going on at present. 

EDMs have been searched for in various systems with different sensitivities 
(Table \ref{edm_limits}). \cite{Khriplovich_1997,Sandars_2001,Semertzidis_2003,NUPECC_2004,
Akesson_2006, Jungmann_2005}.
Distinctively different precision experiments to search for a EDM 
are under way in many different system.A large number of ideas for significant improvements have 
been made public. Still, the electron and the neutron \index{neutron} get the largest attention
of experimental groups, although besides tradition there is little which singles
out these systems. Nevertheless, there is a  large number of efforts in the USA and in Europe 
using different approaches which all have unique promising features.
 
In composed systems, i.e. molecules, atoms or nuclei, 
fundamental particle dipole moments of constituents can be
significantly enhanced\cite{Sandars_2001}.
For the electron significant enhancement factors
are planned to be exploited such as those associated with the large internal 
electric fields in polar molecules\cite{deMille_2002}.

The physical systems investigated fall in six groups (see Fig. {\ref{fig_edm_attack}), i.e.
\begin{itemize}
\item [(i)]   'point' particles (e, $\mu$, $\tau$),
\item [(ii)]  nucleons (n, p),
\item [(iii)] nuclei ($^2$H, $^{223}$Fr, ...),
\item [(iv)] atoms  (Hg, Xe, Tl, Cs, Rn, Ra,...) and
\item [(v)]  molecules (TlF, YbF, PbO,HfF$^+$, ThF$^+$ ...),
\end{itemize}  
where each investigated object has its own particular advantages.
Among the methods employed are
\begin{itemize}
\item [(i)] Classical approaches using optical spectroscopy of atoms and molecules
in cells, as well as atomic and molecular beams or with contained cold neutrons,
\item [(ii)] Modern atomic physics techniques such as atomic and ion traps, fountains and 
                   interference techniques;
\item [(iii)] Innovative approaches involving radioactive species, storage rings, particles in
                   condensed  matter (garnets, superfluid helium) ,  nuclear spin masers \cite{Yoshimi_2005}, 
                  and a few more.
\end{itemize}

From an unbiased point of view there is no preferred system to search for an EDM.
In fact, many different systems need to be examined in order to
be able to extract unambiguous information on the nature of EDMs, because depending
on the underlying yet unknown processes different systems have
in general quite significantly different susceptibility
to acquire an EDM through a particular mechanism (see Figure \ref{fig_edm_sources}).
As a first approach an EDM may be considered an ''intrinsic property'' of
an elementary particle as we know them, because the  
mechanism causing an EDM is not accessible at present. However, an EDM
can also arise from CP-odd forces between the constituents under observation,
e.g. between nucleons in nuclei or between nuclei and
electrons. Such EDMs could be much higher \cite{Liu_2004}  than such
expected for elementary particles originating within the presently popular
New Physics models.

\begin{table}[tbh]
\begin{center}
{\small
\begin{tabular}{l|cccc} 
 Particle & Limit/Measurement & Reference    & SM limit             & possible New Physics \\           
               &        [e\,cm]                &                      & [factor to go]       & [factor to go]  \\ \hline  
e               & $<1.6 \times 10^{-27}$  &  \cite{Regan}         &     $10^{11}$        & $\leq 1$        \\
$\mu$    & $<2.8\times 10^{-19}$   &  \cite{BNL_EDM} & $10^8$     & $\leq 200$      \\ 
$\tau$    & $(-2.2 <d_{\tau}<4.5)\times 10^{-17}$ &  \cite{Belle_tau} & $10^7$     & $\leq 1700$      \\ 
n               & $<2.9 \times 10^{-26}$ &\cite{Harris}     & $10^4$     & $\leq 60$        \\ 
p               & $(-3.7\pm 6.3) \times 10^{-23}$ & \cite{Proton_EDM}                    & $10^7$     & $\leq 10^5$      \\ 
$\Lambda^0$     & $(-3.0\pm 7.4) \times 10^{-17}$ &    \cite{Lambda_EDM} & $10^{11}$ & $10^9$    \\
$\nu_{e,\mu}$   & $<2 \times 10^{-21}$ &  \cite{delAguila} & & \\  
$\nu_{\tau}$    & $<5.2 \times 10^{-17}$ &\cite{NuTau_EDM}                       & & \\    
Hg-atom         & $< 2.1 \times 10^{-28}$ & \cite{Romalis_2001}    & $\leq 10^5$ & various\\ 
\end{tabular}
}
\caption{Some actual limits on EDMs and the improvement factors necessary
in experiments to reach SM predictions. 
For electrons, neutrons and muons the region where speculative models have predicted
a finite value for an EDM can be reached with presently proposed experiments in the near future.
There is a number of new ongoing activities, e.g. in neutral and charged molecules or radioactive atoms, 
which have no reported limit yet. However, they are similarly promising.
\label{edm_limits}} 
\end{center}
\end{table}

\section{Some New Developments in the Field of EDM Searches}
The highly active field of EDM searches has very recently seen a plurality of novel 
ideas. Theoretical work has made very clear that one needs
a number of measurements in various systems, in order to identify
underlying mechanisms of CP-Violation and EDM generation, once the existence
of an EDM has been proven. 
On the experimental side, novel ideas 
led to new activities in systems not investigated so far. Among those
are in particular radioactive atoms and charged particles. The latter had
been excluded by the community for several decades, because of a misinterpretation
of the role of the Lorentz force, which only recently could be convincingly 
cleared up \cite{Farley_2004,Sandars_2001}.

\begin{figure}[tbh]
\begin{center}
\epsfig{file=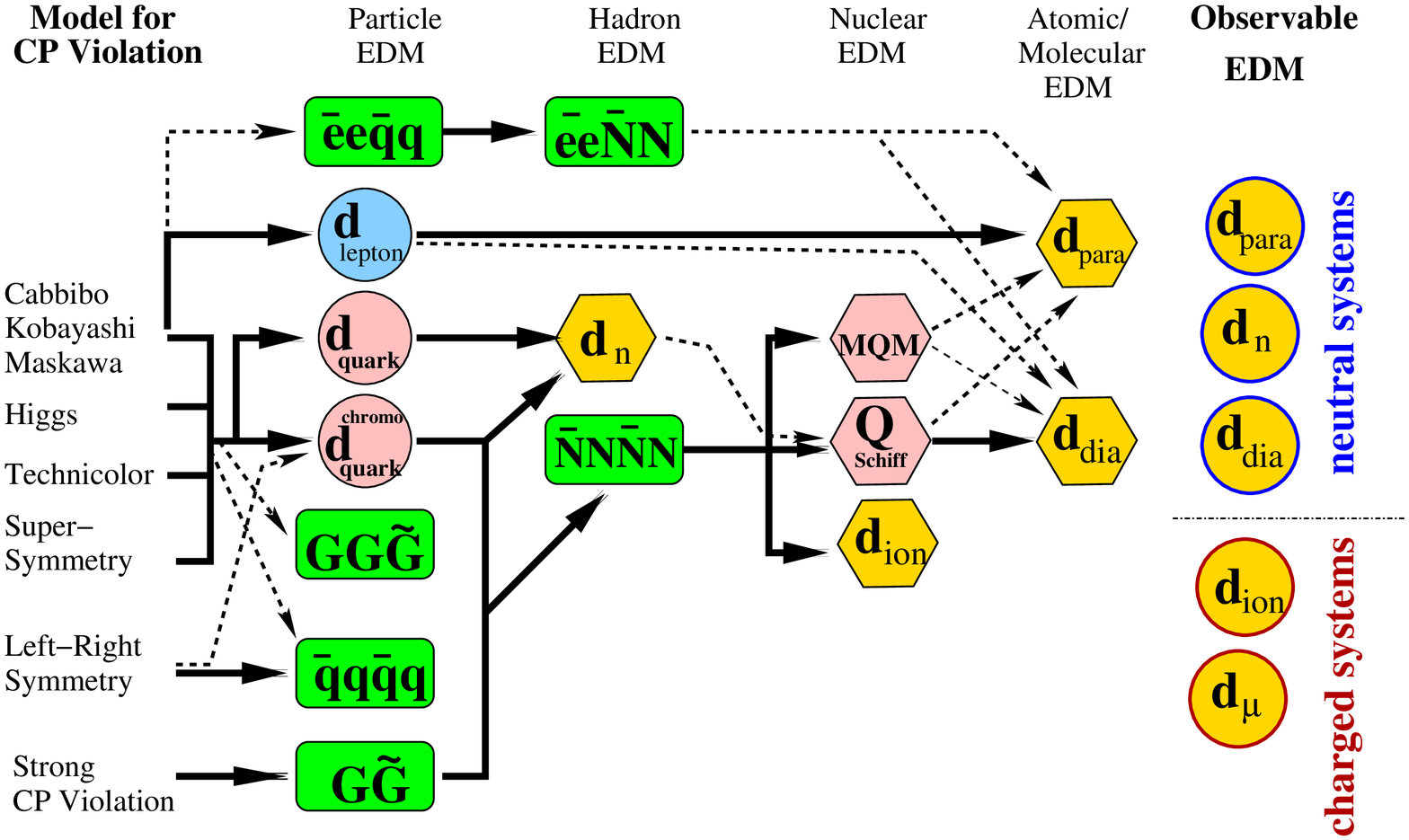,height= 3.5 in}   
\caption{A variety of theoretical speculative models exists 
in which an EDM could be induced through different mechanisms
or a combination of them into fundamental particles and composed 
systems for which an EDM would be experimentally accessible. }
\label{fig_edm_sources}
\end{center}
\end{figure}

\subsection{Radioactive Systems} \index{radioactive atoms}
New facilities around the world make more short-lived radioactive
isotopes available for experiments. Of particular interest for EDM searches is the Ra atom. This atom has   
has as a unique feature in heavy atoms rather close lying  states of opposite parity. 
This accidental almost degeneracy of the $7s7p^3P_1$ and $7s6d^3D_2$  states has led to the prediction of 
a significant enhancement for an  electron EDM \cite{Dzuba_2001} - much
higher than for any other atomic system. An additional advantage of Ra arises from the fact,
that for many of its isotopes their nuclei fall are within in a region 
where (dynamic) octupole deformation occurs, 
which also enhances the effect of a nucleon EDM
substantially, i.e. by some two orders of magnitude \cite{Engel_2004}.
From a technical point of view the Ra atomic levels of interest for an experiment
are well accessible spectroscopically and the isotopes can be produced 
in sufficient quantities in nuclear reactions. 
The advantage of an accelerator based Ra experiment is apparent,
because nuclear EDMs are only possible nuclei with spin
and all Ra isotopes with no-vanishing nuclear spin are relatively 
short-lived \cite{Jungmann_2002}. 
Recently most significant progress towards a Radium EDM \index{Radium}
experiment has been reported from
the Argonne National Laboratory, USA, where successfully a small number of some 10
Ra atoms from an atomic beam source could be stored in a magneto-optical trap (MOT) \cite{Guest_2007}.  Plans were reported to search for a ground state end hence nuclear EDM using a far off resonance optical trap.
Activities towards the excited state EDM of the Ra atoms are under way at KVI \index{KVI}, Groningen, 
The Netherlands \cite{Jungmann_2002,Traykov_2006,Willmann_2006}.

The generic statistical sensitivity of an EDM experiment is given by
\begin{equation}
\delta d = \frac{\hbar}{P\varepsilon T \sqrt{N \tau} E},
\end{equation}
where $P$ is the polarization, $\varepsilon$ the efficiency, $T$ the measurement time, $N$ the flux of particles, $\tau$ the spin coherence time and $E$ the applied electric field. With typical values achievable
in most experiments ($P\approx 1,\varepsilon \approx 1, N= 10^6/s,  \tau \approx 1s, E = 10^5 V/cm$) one gets $\delta d \approx 7 \cdot 10^{-29} e\,cm$ in a day ($T=10^5 s$). Therefore statistics is not
expected to be a serious problem even with exotic and radioactive systems. However, one should
note that systematic effects must be expected the more severe limitations and therefore
they need to be in the center of the attention. 

\subsection{Searches for EDMs in charged Particles}
A very novel idea was introduced recently for measuring an 
EDM of charged particles directly. For such experiments the high
motional electric field is exploited, which charged particles at relativistic speed 
experience in a magnetic storage ring. \index{storage ring}
In such a setup the Schiff theorem can be circumvented
(which had excluded charged particles from experiments due to the
Lorentz force acceleration), because of the non-trivial geometry of the 
problem\cite{Sandars_2001}. With an
additional radial electric field in the storage region the spin precession due to the
magnetic moment anomaly can be compensated, if the 
effective magnetic anomaly $a_{eff}$ is small, i.e. $ a_{eff}<<1$ \cite{Khriplovich_1999}. 

The method was first considered for muons. \index{muon} For longitudinally polarized 
muons injected into the ring an EDM  would express itself 
as a  spin rotation out of the orbital plane.
This can be observed as a time dependent (to first order linear in time) 
change of the above/below  the plane of orbit counting rate ratio. 
For the possible muon beams at the future J-PARC facility in Japan
a sensitivity of $10^{-24}$~e\,cm is expected \cite{Yannis_2003,Farley_2004}. 
Other than for most other EDM searches, in such an experiment the possible muon flux is a major limitation.
For models with nonlinear mass scaling of EDM's such a muon EDM experiment would \index{muon}
already be more sensitive to some certain new physics models
than the present limit on the electron EDM \index{electron}
\cite{Babu_2000}. For certain Left-Right symmetric models a value of $d_{\mu}$
up  to $5\times 10^{-23}$ e\,cm would be possible.
An experiment carried out at a more intense muon source could provide
a significantly more sensitive probe to CP violation in the second 
generation of particles without strangeness.
\footnote{A New Physics (non-SM) contribution $a^{NP}_{\mu}$ 
to the muon magnetic anomaly and a muon EDM $d_{\mu}$
are real and imaginary part of a single complex quantity related through
$d_{\mu}=3 \times 10^{-22} \times (a^{NP}_{\mu}/ (3 \times 10^{-9})) \times \tan \Phi_{CP} e\,cm$
with a yet unknown CP violating phase $\Phi_{CP}$. The problems around the
SM model value for $a_{\mu}$ \cite{Bennett_2006}, which cause  difficulties
for the interpretation of the recent muon g-2 experiment in terms of limits for 
or indications of New Physics,  make a search for $d_{\mu}$ attractive as an important
alternative, as the SM value is negligible for the foreseeable future.}

The deuteron is the simplest known nucleus. Here an EDM
could arise not only from a proton or a neutron EDM, but also
from CP-odd nuclear forces \cite{Hisano_2004}. It was shown very recently \cite{Liu_2004} that
the deuteron \index{deuteron} can be in certain scenarios significantly more sensitive than the
neutron \index{neutron}. The situation is
evident for the case of quark chromo-EDMs, where the EDMs induced into deuteron and neutron are
$ d_{\mathcal{D}}  =  -4.67\, d_{d}^{c}+5.22\, d_{u}^{c}$ and 
$ d_{n}  =  -0.01\, d_{d}^{c}+0.49\, d_{u}^{c}$; i.e. the deutron could have a much higher
sensitivity to quark chromo-EDMs arising from the proton-neutron interaction within the deuteron.
It should be noted that because of its rather small magnetic anomaly
the deuteron is a particularly interesting candidate for a ring EDM experiment
and a proposal with a sensitivity of $10^{-27}$~e\,cm exists \cite{Semertzidis_2004a}.
In this case scattering off a target will be used to observe a spin precession.
As possible sites of an experiment the Brookhaven National Laboratory (BNL),
and KVI are considered.
A further novel approach concerns the search for an electron EDM in charged molecular ions
such as HfF$^+$ and ThF$^+$ where the advantage of easy trapping for ions and the 
strong enhancement factors for electron EDMs in polar molecules \index{molecules} are combined
\cite{Cornell_2007}. The EDM experiments directly using charged particles are still in the exploratory 
and feasibility study phase, however, the progress reports are very encouraging.

\section{T-violation Searches other than EDMs}

There are many more  possibilities to find T-violation besides through searches for EDMs. 
Among the presently ongoing activities certain correlation observables  
in nuclear $\beta$-decays offer excellent opportunities 
to find new sources of CP violation\cite{NUPECC_2004, Akesson_2006,
Herczeg_2001,Jungmann_2004a}. 
In $\beta$-neutrino correlations the 'D'-coefficient\cite{Herczeg_2001}
(for spin polarized nuclei)  offer a high potential to observe new interactions in a region 
of potential New Physics which is less accessible by EDM searches. However,
the 'R'-coefficient\cite{Herczeg_2001} (observation of $\beta$-particle
polarization) would explore the same areas as present EDM searches 
or $\beta$-decay asymmetry measurements.
Such experiments are underway at a number of laboratories worldwide\cite{Jungmann_2004a}.
  
\section{Conclusions}
There is a large field of searches for EDMs on a variety of
systems. They all are well motivated and have unique and robust discovery potentials.
Novel ideas have emerged in the recent past to use yet not studied systems and  
new experimental approaches, which have emerged in the recent past
offer excellent opportunities to complement the more traditional
experimental approaches on neutron-, atom- and electron-EDMs.
Any successful search in the future will have to be complemented by experiments on
other systems in order to pin down eventually the mechanisms 
leading to the observed EDMs. The highest predicted values in beyond
SM speculative theories are well within reach of presently ongoing and planned 
experiments.

\bigskip
This work has been supported by the Dutch Stichting voor Fundamenteel Onderzoek der Materie (FOM) in the framework of the research programme 48 (TRI$\mu$P).\footnote{This conference review of recent developments in the field of EDM searches draws significantly on  earlier write-ups \cite{Jungmann_2005,Jungmann_2004a}.} \index{TRI$\mu$P}

\end{document}